\documentclass[aps,pra,twocolumn,longbibliography]{revtex4-1}

\usepackage{amsmath}
\usepackage{amssymb}
\usepackage{graphicx}
\usepackage{dsfont}
\usepackage[colorlinks = true, linkcolor = blue, citecolor = blue,urlcolor = blue]{hyperref}
\usepackage{natbib}

\newcommand{\subket}[1]{\left|#1\right\rangle_{\text{sub}}}

\newcommand{\sqket}[1]{\left|#1\right\rangle_{\text{sq}}}

\newcommand{\sqbraket}[1]{\left\langle #1 \right\rangle_{\text{sq}}}
\newcommand{\braket}[1]{\left\langle #1 \right\rangle}

\begin{document}

\title{Theory of quantum entanglement and the structure of the two-mode squeezed antiferromagnetic magnon vacuum}
\author{D. Wuhrer}
\email{dennis.wuhrer@uni-konstanz.de}
\affiliation{Fachbereich  Physik,  Universit\"at  Konstanz,  D-78457  Konstanz,  Germany}
\author{N. Rohling}
\email{niklas.rohling@uni-konstanz.de}
\affiliation{Fachbereich  Physik,  Universit\"at  Konstanz,  D-78457  Konstanz,  Germany}
\author{W. Belzig}
\email{wolfgang.belzig@uni-konstanz.de}
\affiliation{Fachbereich  Physik,  Universit\"at  Konstanz,  D-78457  Konstanz,  Germany}

\begin{abstract}
	\noindent  
	Recently investigations of the quantum properties of an antiferromagnet in the spin wave approximation have identified the eigenstates as two-mode squeezed sublattice-magnon states. The uniform magnon states were shown to display a massive sublattice entanglement. Here we extend this idea and study the squeezing properties of all sublattice Fock states throughout the magnetic Brillouin zone.
	We derive the full statistics of the sublattice magnon number with wave number $\vec k$ in the ground state and show that sublattice magnons occur in pairs with opposite wave vectors, hence, resulting in entanglement of both modes. To quantify the degree of entanglement we apply the Duan-Giedke-Cirac-Zoller inequality and show that it can be violated for all modes. The degree of entanglement decrease towards the corners of the Brillouin zone. We relate the entanglement to measurable correlations of components of the N\'eel and the magnetization vectors, thus, allowing to experimentally test the quantum nature of the squeezed vacuum. 
	The distinct $k$-space structure of the entanglement shows that the squeezed vacuum has a nonuniform shape that is revealed through the $\vec k$-dependent correlators for the magnetization and the N\'eel vectors.
\end{abstract}
{\hypersetup{urlcolor = black}
\maketitle
}

\section{INTRODUCTION}
In recent years the desire to increase the computational power, together with the need of electronic devices in a suitable size made it necessary for electric circuits to become smaller and smaller \cite{Moore}. Approaching the size, where quantum effects come into play, it is necessary to come up with new ideas to either limit the influence or use the beneficial effects of the quantum nature. 

One of these ideas is the use  of spin transport to convey information. In metals \cite{SpinTransportMetal} or semiconductors \cite{SpinTransportSemiconductor} spin-transport is achieved by diffusive transport of spin carrying electrons. Interactions between electron spins and magnetic moments in magnetic materials gave then birth to hard drives, based on the giant magnetoresistance \cite{GMR1,GMR2} or tunnel magnetoresistance \cite{TMR}, as a way to store information.

In insulators containing magnetic moments spin-transport can be mediated by collective excitations of spins\cite{MagnonSpintronics, SpinTransportAFMI, SpinTransportFMI}, the so called spin-waves \cite{FreemanDyson,Kittel,NoltingRamakanth}. Similar to light waves that consist of photons, spin waves constitute superpositions of elementary quantized, particle-like excitations, the magnons \cite{BlochMagnons,BrockhouseMagnons}.

Ever since the discovery of 2D materials and the possibility to consistently produce them, they are under extensive investigations \cite{2DMat,Graphen,GraphenEncBN,2DMatMoS,2DMatMoSTrans,2DMat2, 2DMatBN}.  Combining or gating single layers opened the door to tailor materials with desired properties. One could then use different 2D materials and combine them in a van der Waals heterostructure \cite{VdWHeterostructures1, VdWHeterostructures2} to merge their properties or give rise to new ones. Recently magnetic 2D materials joined the zoo of 2D materials \cite{2DMagnetism2, 2DMagnetism3, 2DMagnetic,2DMagnetism1} making it possible to include magnetism into these heterostructures. The so created 2D materials and their spin transport properties are of great interest for future research and spin information processing.

Bose-Einstein condensation (BEC) of magnons was reported in \cite{BoseEinsteinMagnon} as a dynamical, quasi-equlibrium state. Magnons created by a strong drive, are subject to thermalization and are assumed to collapse in the lowest-energy eigenstate resulting in a BEC. The quantum nature of this effects is subject to a debate since alternative classical explanations were made \cite{RuckriegelKopietz}. Another  BEC of magnons occurs in a quantum magnet \cite{BECCondensation}. Here the spin interaction is mapped onto a theory of hard bosonic particles. In the case of an antiferromagnet (AFM) an applied magnetic field $\vec H$ can lower the magnon gap to zero by spin flips. At a critical value $H_{\text{sat}}$ all spins align with $\vec H$. This spin-flipped state can be understood as containing the maximum number of bosons and, hence, as condensate forming the ground state of the AFM.
 
More recently, the field of quantum magnonics has gained some momentum. It was shown that genuine quantum features like squeezing and entanglement exist in several kinds of magnetic insulators \cite{KamraBelzig1, KamraBelzig3}. These quantum features can e.g. been probed by spin-current shot noise measurements \cite{KamraBelzig1}. Even more intriguing is the observation that antiferromagnets constitute a massive source of quantum entanglement between the two sublattices even in the vacuum state \cite{KamraBelzig5}. The investigations so far have concentrated on the uniform magnon mode and motivates our research to investigate quantum entanglement at finite momentum.

To this end, we investigate squeezed magnons as eigenstates of an AFM \cite{SqueezedMagnons, KamraBelzig4, JiSeTserkovnyak}, which utilizes the well-known optical concept of squeezing \cite{gerry_knight_2004} in the domain of antiferromagnetism. Already before, squeezed magnons have shown to be the eigenstates of ferromagnets \cite{KamraBelzig1, KamraBelzig5} in the presence of dipolar interactions and were discussed in the context of ferrimagnets \cite{KamraBelzig3} or, more general, of magnetically ordered  materials \cite{KamraBelzig2}. The fact, that the energy eigenstates of an AFM are already squeezed states and it is not necessary to squeeze them with an external drive is a distinct difference to photon squeezed states. 

Our discussion will be limited to the case of AFMs because we expect the strongest quantum features in that case. The paper is structured as follows. In Sec.~\ref{SecII}, we extend the theory of the squeezed vacuum eigenstate  to all possible wave vectors in the magnetic Brillouin zone. Thereby we show that the $\vec k$-dependent probability $p_{\vec k}$ to find at least one pair of sublattice magnons in the squeezed vacuum is determined by the $\vec k$-dependent squeezing parameter $r_{\vec k}$. Further we find that the sublattice magnons are entangled and show that they violate the Duan-Giedke-Cirac-Zoller (DGCZ) inequality \cite{DuanZoller} which is a clear signature of the quantum nature of the squeezed magnons.
In Sec.~\ref{SecIII}, we show how the $k$-structure of the probabilities $p_{\vec k}$ transfers to the spin-spin correlators and therefore the correlators of magnetization and N\'eel-Vector components.
Finally in Sec.~\ref{SecIV} we will calculate the statistics of the sublattice magnons and its dependence on the ratio between exchange and anisotropy constants.

\section{Theoretical model}
\label{SecII}
We cover a bipartite square lattice AFM with the nearest neighbours of each spin being part of the other sublattice. We do a N\'eel ordered ansatz along the z-axis for both sublattices. The Heisenberg interaction shall act only between nearest neighbours. The Hamiltonian can then be written as
\begin{equation}
	\hat H = - \frac{J}{ \hbar^2}\sum_{<i,j>}\vec S\left(\vec r_i\right)\cdot\vec S\left(\vec r_j\right) - \frac{K}{ \hbar^2}\sum_{i}\left(\hat S^z\left(\vec r_i\right)\right)^2\, , 
\end{equation}
with the strength of the exchange interaction $J < 0$ for AFMs, the strength of the uni-axial anisotropy along the z-axis $K$ and $\vec S\left(\vec r_i\right)$ being the spin operator at  position $\vec r_i$, with components $\hat S^\alpha \left(\vec r_i\right)$, $\alpha \in \{x,y,z\}$.  The division by $\hbar^2$ ensures that both, $J$ and $K$, have the unit of energy.

Applying the standard techniques, with the Holstein-Primakoff transformation \cite{HolsteinPrimakoff} in linear approximation, the Hamiltonian becomes
\begin{equation}
\hat{H} = \sum_{\vec k}A_{\vec k} \left(\hat{a}_{\vec k}^\dagger \hat{a}_{\vec k} + \hat{b}_{\vec k}^\dagger \hat{b}_{\vec k}\right) + C_{\vec k}\left(\hat{a}_{\vec k}^\dagger \hat{b}_{-\vec k}^\dagger + \hat{a}_{\vec k} \hat{b}_{-\vec k}\right)\, .
\end{equation}
To use the linear spin wave approximation in the Holstein Primakoff transformation, which is an expansion of $\sqrt{1 - \hat n/2S}$ in terms of small $\hat n/2S$, we have to exclude the case of $S = 1/2$. Further, we emphasise to pay attention to a possible break down of the approximation in case of low spin lengths/huge magnon numbers. 

In the Hamiltonian, the different interactions are covered in the two functions, which are given as $A_{\vec k} = A =S\left( 2K - Jz\right)$, $C_{\vec k} = - JSz\gamma_{\vec k}$ and $\gamma_{\vec k} = \sum_{\vec \delta} e^{\mathrm{i} \vec{\delta} \vec k}/z$. Here, $z$ is the number of nearest neighbours at each lattice site, which depends on the dimensionality of our system, and $\vec \delta$ are the connection vectors to these nearest neighbours, $\hat a_{\vec k}$ ($\hat{b}_{\vec k}$) is the annihilation operator of a magnon with wave vector $\vec k$ in the A (B) sublattice. We set the lattice constant to $a = 1$. The diagonal components $A_{\vec k}$ can be $\vec k$-dependent if one does not limit to nearest-neighbour interaction. Terms arising during the manipulation of the Hamiltonian and which result in a constant energy shift are neglected. 

We then perform a Bogoliubov transformation connecting the sublattice operators with new creation and annihilation operators $\hat \alpha$ and $\hat \beta$ via 
\begin{equation}
		\left(\begin{array}{c}
	\hat{a}_{\vec k}\\
	\hat{b}_{-\vec k}^\dagger
	\end{array}\right)  = \left(\begin{array}{cc}
	u_{\vec k}& v_{\vec k}\\
	v_{\vec k}^* & u^\ast_{\vec k}
	\end{array}\right)\left(\begin{array}{c}
	\hat{\alpha}_{\vec k}\\
	\hat{\beta}_{-\vec k}^\dagger
	\end{array}\right) \, .
	\label{eq:Bogoliubov}
\end{equation}
The Bogoliubov transformation will diagonalize the Hamiltonian, if we demand $\hat \alpha$ and $\hat \beta$ to be bosonic operators and choose the matrix elements such, that they satisfy 
\begin{figure}
	\begin{center}
			\includegraphics[width = \columnwidth]{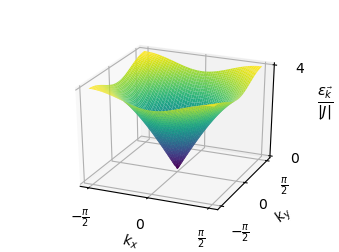}
	\end{center}
	\caption{Energy dispersion for a two dimensional square lattice antiferromagnet, with $|J|/K = 10^4$.}
	\label{fig:0}
\end{figure} 
\begin{equation}
	\begin{split}
	 u_{\vec k} &=\phantom{-} \sqrt{\frac{A + \varepsilon_{\vec k}}{2\varepsilon_{\vec k}}}\, ,\\
	 v_{\vec k} &= -\sqrt{\frac{A - \varepsilon_{\vec k}}{2\varepsilon_{\vec k}}}\, ,
	\end{split}
	\qquad \varepsilon_{\vec k} = \sqrt{A^2 - C_{\vec k}^2}\, ,
	\label{eq:DefUV}
\end{equation}
where $\varepsilon_{\vec k}$ is the energy of one $\hat \alpha$ ($\hat \beta$) magnon with wave vector $\vec k$.
\subsection{Magnon squeezing}
We introduce the two-mode squeezing operator \cite{gerry_knight_2004} 
\begin{equation}
\hat{S}_2\left(r_{\vec k}\right) = \exp{\left[r_{\vec k}\left(\hat a_{\vec k} \hat b_{-\vec k}-\hat a_{\vec k}^\dagger  \hat b_{-\vec k}^\dagger\right)\right]}\, ,
\label{eq:DefSquOp}
\end{equation}
with the positive squeezing parameter $r_{\vec k}$, which can be connected to the Bogoliubov transformation via $\cosh r_{\vec k} = u_{\vec k}$ and $\sinh r_{\vec k} = -v_{\vec k}$. Applying $\hat S(r_{\vec k})$ to the sublattice vacuum state $\subket{0}$ creates an entangled state of both modes $\hat a_{\vec k}$ and $\hat b_{-\vec k}$. $p_{\vec k} = \tanh{(r_{\vec k})}^{2}\in [0,1]$ is the probability to find  at least one pair of magnons, consisting of one magnon with wave vector $\vec k$ in sublattice A and one magnon with wavevector $- \vec k$ in sublattice B. The application of the squeezing operator results in a squeezing of the quadratures of superpositions of both modes. In the picture of two harmonic oscillators this squeezing can be imagined as a squeezing of the relative and center of mass coordinates.

The squeezing parameter determines the degree of squeezing in our system and is determined by $|J|/K$. By changing the anisotropy strength one can then tune the degree of squeezing in our system, similar to the squeezing in a ferromagnet \cite{JiSeTserkovnyak}. 

A straightforward calculation shows, that one can write the new magnon operators as 
\begin{equation}
\hat \alpha_{\vec k} =\hat S_2\left(r_{\vec k}\right) \hat a_{\vec k}\hat S_2\left(r_{\vec k}\right)^{-1},\quad\hat \beta_{\vec k} = \hat S_2\left(r_{\vec k}\right) \hat b_{\vec k}\hat S_2\left(r_{\vec k}\right)^{-1}\, ,
\end{equation} 
which is why we will call them "squeezed magnons" from now on.

If we denote the ground state, or vacuum state of our system as $\sqket{0}$ and the N\'eel state, or sublattice vacuum state, as $\subket{0}$ and apply $\hat \alpha_{\vec k} \sqket{0} = 0$, one can show, that we can connect both vacua via \cite{SqueezedMagnons} 
\begin{equation}
	\sqket{0}= \prod_{\vec k}\hat S_2\left(r_{\vec k}\right)\subket{0}\, ,
\end{equation} 
resulting in the squeezed vacuum being an entangled, squeezed state of sublattice modes.

By applying the squeezed annihilation operators, expressed through the Bogoliubov transformation (Eq.~(\ref{eq:Bogoliubov})), onto the squeezed vacuum a straightforward calculation shows
\begin{equation}
	\sqket{0} = \sum_{\vec n = \vec 0}^{\infty}\prod_{\vec k}\left(\frac{\left(-\tanh r_{\vec k}\right)^{n_{\vec k}}}{\cosh r_{\vec k}}\right)\subket{\vec n, \vec m(\vec n)}\, .
	\label{eq:expansion}
\end{equation}
Here the components of $ \vec n$ ($\vec m$) are the occupation numbers of the different sublattice modes in the A (B) sublattice, with $n_{\vec k}$ ($m_{\vec k}$) being the number of magnons in the mode with wave vector $\vec k$,  $m_{\vec k}(\vec n) = n_{-\vec k}$ and 
\begin{align}
    \begin{split}
    \subket{\vec n, \vec m(\vec n)} &=\bigotimes_{\vec k} \subket{n_{\vec k}, m_{\vec k}(\vec n)} =\\
    &= \bigotimes_{\vec k}\left(\subket{n_{\vec k} }^A \otimes \subket{n_{-\vec k}}^B\right)\, ,
    \end{split}
\end{align}
where $\subket{\ldots}^{A/B}$ is a state in sublattice A (B).
We see, that we only have states with the same number of magnons in the A and B sublattice for opposite wave vectors. This will guarantee us an overall vanishing momentum in each mode. Furthermore, one can show that the expectation value of the spin of a magnon in the A and a magnon in the B sublattice are antiparallel, resulting in an overall vanishing expectation value for the spin operator. 

If we want a spin-up magnon of wave vector $\vec k$ in the squeezed space, we have to act with a creation operator $\hat \beta_{\vec k}^\dagger$ onto the squeezed vacuum. Using the expression for $\hat{\beta}_{\vec k}$ via the squeezing operator and the expansion of the squeezed vacuum, we get for the expansion of the one-magnon state
\begin{equation}
	\begin{split}
		\sqket{\uparrow, \vec k}& = \sum_{\vec n = \vec 0}^{\infty}\left[\prod_{\vec k^{'}\neq \vec k}\left(\frac{\left(-\tanh r_{\vec k^{'}}\right)^{n_{\vec k^{'}}}}{\cosh r_{\vec k^{'}}}\right)\right.\\
		&\left.\times \frac{\sqrt{n_{\vec k}+1}\left(-\tanh r_{\vec k}\right)^{n_{\vec k}}}{\cosh^2 r_{\vec k}}\subket{\vec n, \vec m(\vec n) + \vec e_{\vec k}}\right]\, .
	\end{split}
	\label{eq:expansion2}
\end{equation} 
In contrast to the squeezed vacuum state, we can see, that only sublattice states contribute, if they contain one magnon with wave vector $\vec k$ more in the B sublattice, than magnons with wave vector $-\vec  k$ in the A sublattice. The only probability amplitude, which is changed in the squeezed magnon state with wave vector $\vec k$, is the probability amplitude for the $\vec k $ sublattice modes. 

Applying $\hat \alpha_{\vec k}^\dagger$ onto the squeezed vacuum would lead to a squeezed spin down magnon with the same probability amplitudes for the sublattice states, but we would get one magnon with wave vector $\vec k$ more in the A sublattice than $- \vec k$ magnons in the B sublattice. 
\begin{figure}
	\begin{center}
		\includegraphics[width = 8.6cm]{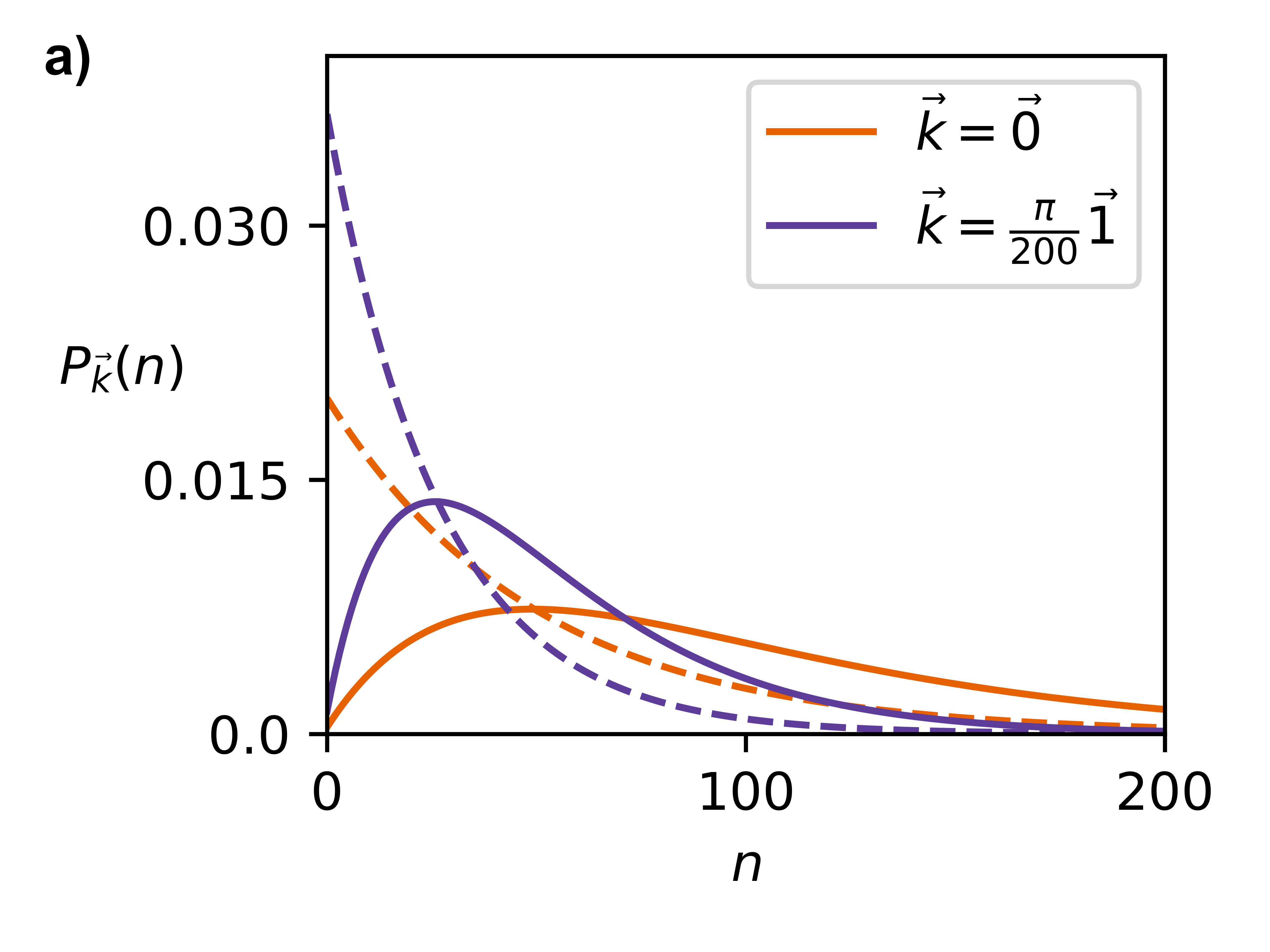}
	\end{center}
	\begin{center}
			\includegraphics[width = 8.6cm]{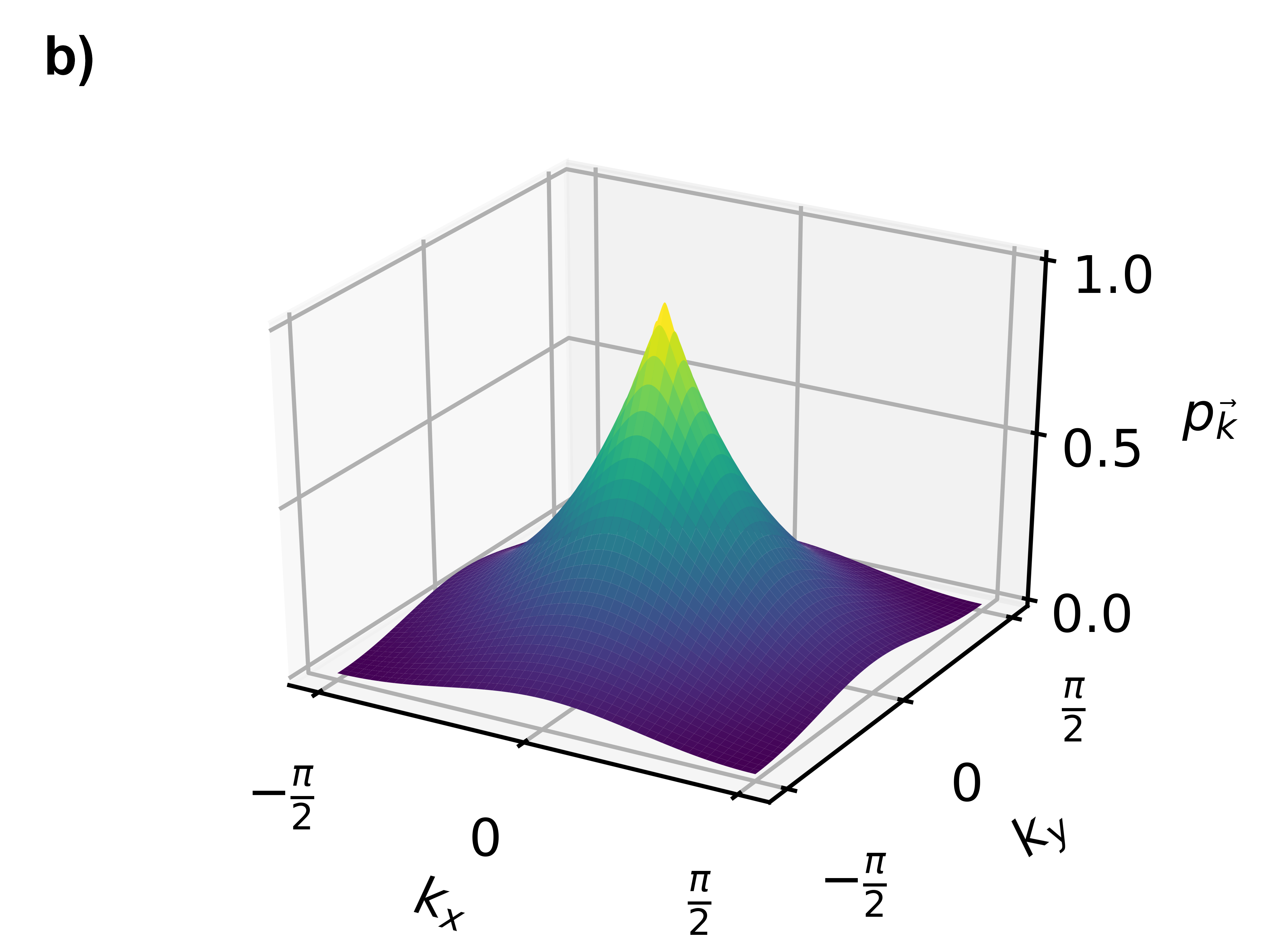}
	\end{center}
	\caption{Both figures were made for $|J|/K = 10^4$. 
	a) Probability $P_{\vec k}(n)$ to find $n$ magnons in sublattice A/B with wave vector $\vec k = \vec 0$ (orange) and $\vec k = \frac{\pi}{200}\vec 1$ (purple) in the squeezed vacuum state (dashed) and the probability $Q_{\vec k}(n)$ to find $n$ magnons in sublattice A/B and $n+1$ in B/A  in the squeezed magnon state (solid). Here $\vec 1 =  (1,1)^T$. b) Probability $p_{\vec k}$ to have at least one magnon in each sublattice. We see a sharp peak at $\vec k = \vec 0$, with $p_{\vec 0} \approx 0.6$, implying, that the probability to find any number of magnons different from $0$ for the other wave vectors is very small. }
	\label{fig:1}
\end{figure} 
In Eq.~(\ref{eq:expansion}) and Eq.~(\ref{eq:expansion2}) we can see that the probability amplitude and therefore the probability takes the same form for all wave vectors, only differing through $r_{\vec k}$.

For the squeezed vacuum the probability to find $n$ pairs of magnons, consisting of one magnon in sublattice A with wave vector $\vec k$  and one magnon in sublattice B with wave vector $-\vec k$, can be given as 
\begin{equation}
	P_{\vec k}(n) = \frac{\tanh{\left(r_{\vec k}\right)^{2n}}}{\cosh{\left(r_{\vec k}\right)^2}} = \left(1 - p_{\vec k}\right)p_{\vec k}^{n}\, .
\end{equation}
The probability for a squeezed spin up (down) magnon with wave vector $\vec k$ containing $n+1$ sublattice magnons of wave vector $\vec k$ in sublattice B (A) and  $n$ sublattice magnons of wave vector $-\vec k$ in sublattice A (B) can also be expressed through $p_{\vec k}$ by
\begin{equation}
	Q_{\vec k} (n) = \left(1 - p_{\vec k}\right)^2 (n+1) p_{\vec k}^{n}\, .
\end{equation} 

Figure \ref{fig:1}a) shows the dependence of $P_{\vec k}(n)$ and $Q_{\vec k}(n)$ on $n$ for the two wave vectors $\vec k = \vec 0$ and $\vec k = \frac \pi {200} (1,1)^\top$. While in the vacuum state each $P_{\vec k}(n)$ falls off exponentially with the maximum at $n = 0$, this changes for $Q_{\vec k}(n)$, in the case of a squeezed magnon, where the function is nonmonotonic, with a maximum at some $n$ different from zero. This behaviour is similar for all $\vec k$, but  the closer to the edge of the Brillouin zone $\vec k$ is, the steeper the probability falls off around $n = 0$ for the vacuum state and the closer is the maximum of $Q_{\vec k}(n )$ to zero. This behaviour can be seen in Fig.~\ref{fig:1}b), which shows $p_{\vec k}$, i.e. the probability to find at least one pair of magnons in the sublattices with corresponding wave vectors. As $p_{\vec k}$ goes towards zero for large wave vectors the expectation value should be centered more around zero magnons, which is confirmed by Fig.~\ref{fig:1}a). 

It is also worth noting the experimental verification of squeezing in the AFMs $\text{MnF}_2$ \cite{MagnonSqueezingExp1} and $\text{FeF}_2$ \cite{MagnonSqueezingExp2} of Zhao et al. Via pump-probe experiments they drove the AFMs into a squeezed magnon state in which the probe pulse detects the two magnon squeezed state. We have to point out the important difference of this light-driven squeezed state to the squeezed eigenstates of our investigated system that exist without an external drive. Further we note, using the formalism of this paper, their ground state is already a squeezed state of magnons and the light induced squeezing describes a squeezing of the already squeezed magnons. The experiments are nevertheless a proof of the existence of magnon squeezing. 

Further, an application of magnon squeezing was suggested by Skogvoll et al. \cite{ApplicationMagnonSq}. They predicted an entanglement of three spin qubits simultaneously coupled to the same ferromagnet. Due to the composite nature of squeezed magnons, which can be seen in Eq.~\eqref{eq:expansion2}, it is possible to excite all three spin qubits by a single squeezed magnon. Resulting in a three-qubit entangled state, namely a Greenberger-Horne-Zeilinger state.

\subsection{DGCZ inequality}
\begin{figure}
	\begin{center}
		\includegraphics[width = 8.6cm]{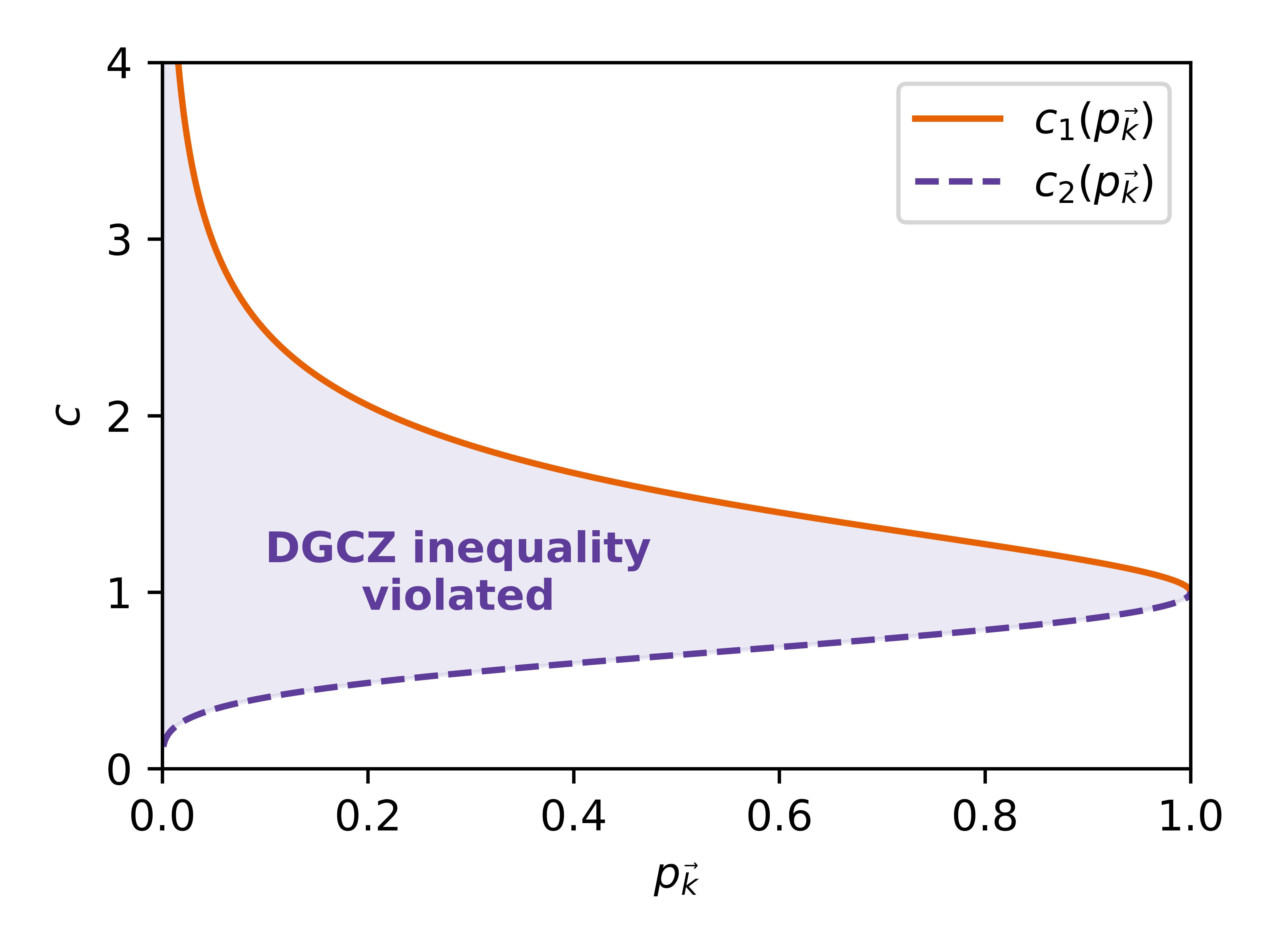}
	\end{center}
	\caption{Values for $c$, depending on $p_{\vec k}$, which violate the DGCZ inequality (purple area). The orange/solid (purple/dashed) line satisfies the equality and correspond to solution $c_{1}$ ($c_2$).}
	\label{fig:2}
\end{figure}

From the full expansion of the squeezed vacuum and one-magnon state we can see, that we cannot separate the $\hat a_{\vec k}$ and $\hat b_{-\vec k}$ modes resulting in an entanglement of both modes. This entanglement is confirmed by the DGCZ inequality \cite{DuanZoller}, which states, that for any separable quantum state the total variance of a pair of operators 
\begin{equation}
	\begin{array}{c}
		\hat u = \left|c\right| \hat x_1 +  \frac 1 c \hat x_2\, ,\\
		\\
		\hat v = \left|c\right| \hat p_1 -  \frac 1 {c} \hat p_2\, ,
	\end{array} \qquad \begin{array}{c}\left[\hat x_i, \hat p_j\right]  = \mathrm i \delta_{i,j} \, ,\\
	\\
	c \in \mathds{R}\backslash\{0\}
	\end{array}
	\label{DefDuan}
\end{equation}
satisfies the inequality
\begin{equation}
	\braket{\left(\Delta \hat u\right)^2} + \braket{\left(\Delta \hat v\right)^2}  \geq c^2 + \frac 1 {c^2} \,.
\end{equation}
If we use the sublattice annihilation and creation operators to define 
\begin{equation}
	\begin{array}{c}
		\hat x_1 = \frac 1 {\sqrt 2} \left( \hat a_{\vec k} + \hat a_{\vec k}^\dagger\right)\, ,\\
		\\
		\hat x_2 = \frac 1 {\sqrt 2} \left( \hat b_{\vec k'} + \hat b_{\vec k'}^\dagger\right)\, ,
	\end{array} \qquad \begin{array}{c}
	\hat p_1 = \frac 1 {\sqrt 2 \mathrm i} \left( \hat a_{\vec k} - \hat a_{\vec k}^\dagger\right)\, ,\\
	\\
	\hat p_2 = \frac 1 {\sqrt 2\mathrm i} \left( \hat b_{\vec k'} - \hat b_{\vec k'}^\dagger\right)\, ,
\end{array}
\end{equation}
then $ [\hat x_i, \hat p_j]  = \mathrm i \delta_{i,j} $ is obviously fulfilled.  If we regard the squeezed vacuum state, indicated through the subscript \textquotesingle sq\textquotesingle\,  further on, we can show for $\vec k' \neq -\vec k$
\begin{equation}
	\begin{split}
	\sqbraket{\left(\Delta \hat u\right)^2}+ \sqbraket{\left(\Delta \hat v\right)^2}&= c^2\frac{1 + p_{\vec k}}{1- p_{\vec k}} + \frac 1 {c^2}\frac{1 + p_{\vec k'}}{1- p_{\vec k'}}  \\
	&\geq c^2 + \frac 1 {c^2}
	\end{split}
\end{equation}
the inequality is always fulfilled. For modes $\vec k' = - \vec k$ we get 
\begin{equation}
	\begin{split}
		\sqbraket{\left(\Delta \hat u\right)^2} + \sqbraket{\left(\Delta \hat v\right)^2}= &\left(c^2 + \frac 1 {c^2} \right)\frac{1 + p_{\vec k}}{1- p_{\vec k}}\\&- 4 \frac{\left|c\right|}{c}\frac{\sqrt{p_{\vec k}}}{1-p_{\vec k}}\, .
	\end{split}
\end{equation}
If we use this equation and demand the DGCZ inequality to be an equality, we can solve this for $c$,
\begin{equation}
	c_{1/2} = \sqrt{\frac 1 {\sqrt{p_{\vec k}}} \left(1 \pm \sqrt{1 - p_{\vec k}}\right)}\, .
\end{equation}
Due to the absolute value of $c$, which plays a role in the DGCZ equation, only the positive solutions $c_{1/2}$ solve the problem at hand. All values for $c$ enclosed by these both branches, which we can see in Fig.~\ref{fig:2}, violate the inequality. From this follows, that all modes $\hat a_{\vec k}$ and $\hat b_{-\vec k}$ are entangled, which supports our earlier claim.
\section{Correlators}

\begin{figure*}
	\begin{minipage}{.5\textwidth}
		\begin{center}
			\includegraphics[width = 8.6cm]{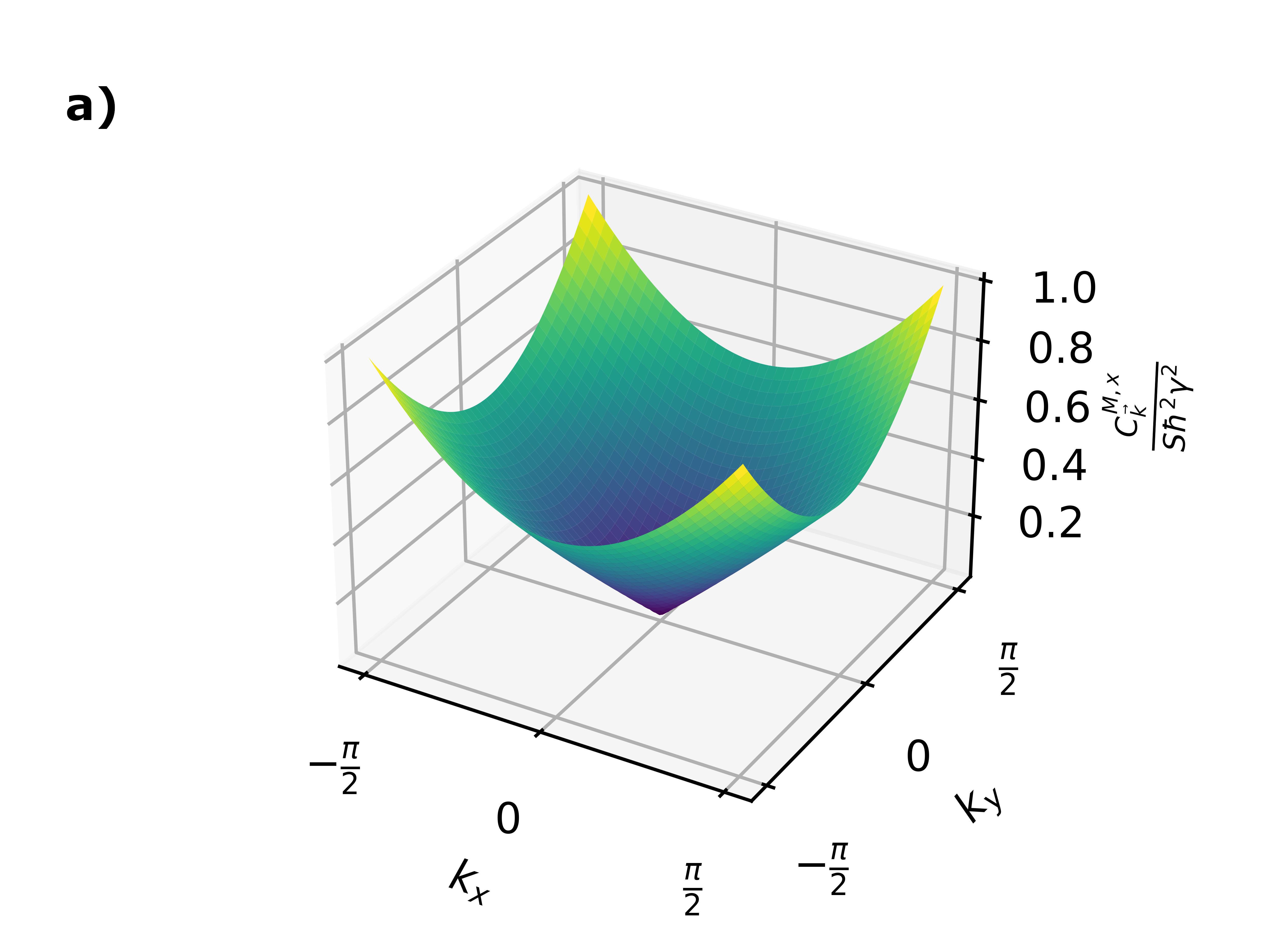}
		\end{center}
	\end{minipage}%
	\begin{minipage}{.5\textwidth}
		\begin{center}
			\includegraphics[width = 8.6cm]{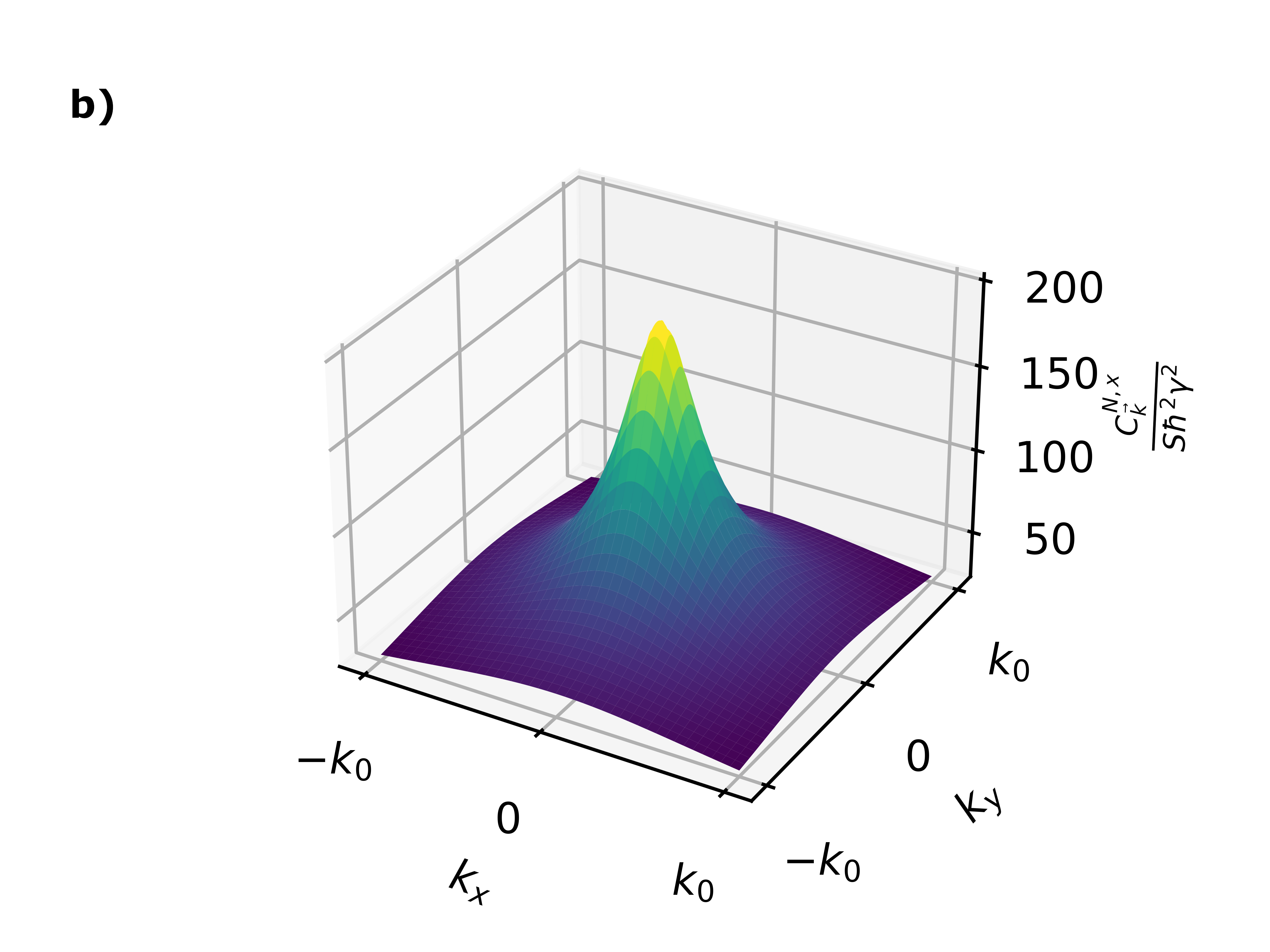}
		\end{center}
	\end{minipage}
	\caption{Correlators of the x-components of the magnetization (left) and N\'eel vector (right) in the $k$ space for $\left|J\right|/K = 10^{4}$. The $k_x$ and $k_y$ values are limited to $k_0 = 0.076$ in picture b) to keep the focus on points with $C_{\vec k}^{N,x} > 1$. For the value of $k_0$ see App.~\ref{SecA2}.}
	\label{fig:3}
\end{figure*} 
\label{SecIII}
The $k$-space structure of $p_{\vec k}$ transfers to the $k$-space structure of the  spin-spin correlation functions and therefore to the correlators of the magnetization and N\'eel vector components, which includes the uncertainty of each variable. The magnetization of an AFM is defined as the sum of the sublattice magnetizations, while the N\'eel vector is the difference of them. At a point $\vec r$ at the lattice both of them are defined as
\begin{align}
	\vec{M}\left(\vec r \right)&= \gamma\left\{\begin{array}{cc}\phantom{-} \vec S^A\left(\vec r\right)&\vec r \in A  \\ \phantom{-}\vec S^B\left(\vec r\right) &\vec r \in B  \end{array}\right. \, , \\
		\vec{N}\left(\vec r \right)&= \gamma\left\{\begin{array}{cc} \phantom{-} \vec S^A\left(\vec r\right)&\vec r \in A  \\  -\vec S^B\left(\vec r\right)&\vec r \in B  \end{array}\right. \,.
\end{align}
Here $\gamma$ is the gyromagnetic ratio. In $\vec k$-space they are given by
\begin{equation}
	\vec M_{\vec k} = \gamma \left(\vec S^A_{\vec k} + \vec S^B_{\vec k}\right)\, , \qquad \vec N_{\vec k} = \gamma \left(\vec S^A_{\vec k} - \vec S^B_{\vec k}\right)\, .
	\label{eq:MagNelk}
\end{equation}
As before $\vec S^A_{\vec k}$ is connected to the sublattice creation and annihilation operators  via the linearised Holstein-Primakoff transformation. This connection yields for the x,y-components of the magnetization and the N\'eel vector in terms of creation and annihilation operators  
\begin{align}
    M_{\vec k}^{x} &= \gamma \hbar \frac{\sqrt{2 S}}{2}\left(\hat{a}_{\vec k} + \hat{a}_{-\vec k}^\dagger + \hat{b}_{\vec k} + \hat{b}_{-\vec k}^\dagger \right)\, ,\\
    M_{\vec k}^{y} &= \gamma \hbar \frac{\sqrt{2 S}}{2\mathrm  i}\left(\hat{a}_{\vec k} - \hat{a}_{-\vec k}^\dagger - \hat{b}_{\vec k} + \hat{b}_{-\vec k}^\dagger \right)\, ,\\
    N_{\vec k}^{x} &= \gamma \hbar \frac{\sqrt{2 S}}{2}\left(\hat{a}_{\vec k} + \hat{a}_{-\vec k}^\dagger - \hat{b}_{\vec k} - \hat{b}_{-\vec k}^\dagger \right)\, ,\\
    N_{\vec k}^{y} &= \gamma \hbar \frac{\sqrt{2 S}}{2 \mathrm  i}\left(\hat{a}_{\vec k} - \hat{a}_{-\vec k}^\dagger + \hat{b}_{\vec k} - \hat{b}_{-\vec k}^\dagger \right)\, .
\end{align}
The correlator of two operators $\vec A$ and $\vec B$ with components $\hat A^\alpha$ and $\hat B^\beta$  ($\alpha, \beta \in \{x,y,z,+,-\}$) at point $\vec k$ in the $k$-space space is defined as
\begin{align}
    C_{\vec k}^{A,B|\alpha,\beta} &= \braket{\hat A_{\vec k}^{\alpha}\hat B_{\vec k}^{\beta\dagger} } - \braket{\hat A_{\vec k}^\alpha} \braket{\hat B_{\vec k}^{\beta\dagger} }\, .
\end{align}
Due to translation symmetry, same $\vec k$ values in both operators are the only points in the $k$-space at which the correlators for magnetization and N\'eel vector components will be non-zero. Keeping in Mind, that $\hat M_{\vec k}^{x\dagger} = \hat M_{-\vec k}^x$ and similar for other components and the N\'eel vector, the correlators of the same vector component become 
\begin{align}
	C_{\vec k}^{M,x}&=C_{\vec k}^{M,y} =  S \hbar^2 \gamma^2\frac{1 - \sqrt{p_{\vec k}}}{1 + \sqrt{p_{\vec k}}} =  S \hbar^2 \gamma^2e^{-2\left|r_{\vec k}\right|}\, ,\label{eq:CorrMag}\\
	C_{\vec k}^{N,x}&=C_{\vec k}^{N,y} =  S \hbar^2 \gamma^2\frac{1 + \sqrt{p_{\vec k}}}{1 - \sqrt{p_{\vec k}}} =   S \hbar^2 \gamma^2 e^{+2\left|r_{\vec k}\right|}\, .
	\label{eq:CorrNel}
\end{align} 

The dependency of the correlation functions on the wave vector components $k_x$ and $k_y$ can be seen in Fig.~\ref{fig:3}. The strongest correlation for the magnetization is at the edge of the Brillouin zone and the weakest at the center $\vec k = 0$, which is the opposite as in the case of the N\'eel vector. For the N\'eel vector the strongest correlations lie in a circle with radius $k_0 = \sqrt{2K/|J|}$ around the origin. This value for $|\vec k|$ determines the point at which the energy dispersion relation can be assumed to be linear, as shown in \ref{SecA2}.

Using the results from section \ref{SecII} one can connect them to the correlators of the magnetization and N\'eel vector components by choosing $c = \pm1$, $\vec k' = - \vec k$ and multiplying $\hat x_i$ and $\hat p_i$ with $\sqrt{2S}\hbar \gamma$ in  Eq.~\ref{DefDuan}, which results in
\begin{equation}
    \sqbraket{\left(\Delta \hat u\right)^2} + \sqbraket{\left(\Delta \hat v\right)^2} = \left\{\begin{split}
      &4\sqbraket{M_{k}^x M_{k}^{x\dagger}}&&(+)\\
      \\
      &4\sqbraket{N_{k}^y N_{k}^{y\dagger}}& &(-)\,,
    \end{split}\right.
\end{equation}
where we used $C^{N/M, x}_{\vec k}= C^{N/M, y}_{\vec k}$ (Eq.~(\ref{eq:CorrMag})+(\ref{eq:CorrNel})). This is in accordance with the choice of $c$ and the earlier calculated violation of the inequality. This means, for $c = 1$ the inequality is always violated, resulting in $\sqbraket{M_{k}^x M_{k}^{x\dagger}}\leq S\hbar^2\gamma^2$, as can be seen from Eq.~(\ref{eq:CorrMag}). For $c = -1$ the inequality is always fulfilled, resulting in $\sqbraket{N_{k}^y N_{k}^{y\dagger}}\geq S\hbar^2\gamma^2$, as can be seen from Eq.~(\ref{eq:CorrNel}). From these inequalities we can see, that the squeezing property of the squeezed vacuum results in a squeezing of the correlators  of the $x/y$-components of the magnetization and N\'eel vector. While one becomes smaller, the other becomes bigger, but the product of both of stays always the same
\begin{equation}
    \sqbraket{M_{k}^x M_{k}^{x\dagger}}  \sqbraket{N_{k}^y N_{k}^{y\dagger}} = \left(S\hbar^2 \gamma^2 \right)^2\, .
\end{equation}
It is important to note here, that the product of the uncertainty of the same spatial components of the N\'eel vector and the magnetization results also in $(S\hbar^2 \gamma^2 )^2$. While the above relation seems to indicate a squeezing between, for example, for $M^x_{\vec k}$ and $N_{\vec k}^x$, we want to emphasize, that $[M^\alpha_{\vec k}, N^\beta_{- \vec k}] = 2\mathrm i S \hbar^2 \gamma^2( 1 - \delta_{\alpha, \beta})$, $\alpha, \beta \in \{x,y\}$. Regarding the uncertainty principle \cite{UncertaintyPrinciple}, the commutator yields, that while for different spatial components the uncertainty is minimal, this is not true for the same spatial components.

For the correlators in z-direction, if we only regard the terms up to order two in the creation and annihilation operators, then the correlators between the $z$-components vanish. The correlation function for different components of the same vectors, e.g. $C_{\vec k}^{M|x,y}$ or $C_{\vec k}^{M|z,x}$, also vanish, as well as the correlators of the same component of different vectors. 

The only other, non-vanishing correlator is between the $x$- and $y$-components of the magnetization and the N\'eel vector, which is given as 
\begin{equation}
	C_{\vec k}^{M,N|x,y}= \mathrm i \gamma^2 S \hbar^2  \,.
\end{equation}
This is a purely imaginary result. However, an expectation value of a measurable quantity must be real. We want to note, that such a measurable quantity is expressed by the anticommutator of $M_{\vec k}^x$ and $N_{\vec k}^y$, which is a Hermitian operator and therefore results in a real expectation value, which is equal to $0$.

The final result of this section is the relation between the correlators of the $x$-component of the magnetization modes an the $y$-componenta of the N\'eel vector modes. It always holds
\begin{align}
    C_{\vec k}^{M,x} \leq   C_{\vec k}^{N,y}\, .
\end{align}
The relation between the correlation functions is of utmost importance as measurement and comparison of the different correlators can be used to determine a clear experimental signature of a squeezed magnonic ground state.

One experimental method to investigate magnetic structures is the well established neutron scattering \cite{NeutronScattering1,NeutronScattering2}. Here the dynamic spin correlation function plays an essential role in determining the differential scattering cross-section. The so-called static spin correlation functions, which we simply call spin-spin correlation functions and which are given in Eq.~(\eqref{eq:CorrMag}) and calculated in appendix \ref{SecA3}, are connected to the dynamic spin correlation function, by integrating its energy dependent, dynamic part out. The dynamic spin correlation function can than be factorised into a frequency (energy) and polarization dependent part and the static spin correlation function. This offers a direct experimental approach to the result calculated in our paper, if one knows the material parameters $J$ and $K$. 

We also want to mention the recent measurement done by Bossini et al.\cite{ExpBossini}. They perform a pump probe experiment at the cubic (tetragonal) lattice AFM $\text{KNiF}_3$ ($\text{K}_2\text{NiF}_4$), from which especially the first one is of interest for our model. They measure the Kerr-rotation of a probe pulse after it is reflected from the AFM surface. This Kerr rotation is directly connected to the spin-spin correlators of spins in the different sublattices. There, our results comes into play. While they themselves performed calculations to determine the correlator, they did not interpret their theory in terms of magnon squeezing, while it is, as we showed, strongly connected. Therefore, we believe that connecting the Kerr-rotation to the spin-spin correlators could be a way to determine the squeezing in the system.

\section{Magnon number}
\label{SecIV}
Finally, we want to investigate the total number of magnons per sublattice in the squeezed vacuum of our system. Therefore, we calculate the expectation value and variance of $\hat n = \sum_{\vec k} \hat a_{\vec k}^\dagger \hat a_{\vec k}$, which is the number operator of all sublattice magnons of sublattice A. The number of magnons in sublattice B is equal in the squeezed vacuum. We use the expansion of the squeezed vacuum in sublattice states in Eq.~(\ref{eq:expansion}) and obtain the expectation value
\begin{equation}
	\sqbraket{\hat n} = \sum_{\vec k } \frac{p_{\vec k}}{1 - p_{\vec k}} = \sum_{\vec k } \sinh^2(r_{\vec k}) 
	\label{eq:expecMagNumb}
\end{equation}
and the variance 
\begin{equation}
\begin{split}
    &\sqbraket{\left(\Delta \hat n\right)^2} = \sum_{\vec k }\sqbraket{\hat n_{\vec k}}\left(\sqbraket{\hat n_{\vec k}} +1\right)  =\\ &= \sum_{\vec k } \frac{p_{\vec k}}{\left(1 - p_{\vec k}\right)^2}= \sum_{\vec k }\cosh^2(r_{\vec k})\sinh^2(r_{\vec k}) \, .
    \label{eq:VarianceN}
\end{split}
\end{equation}
 We can see from Eq.~\eqref{eq:expecMagNumb} that with rising number of possible magnon modes, i.e. with increasing system size, the expectation value as well as the variance will rise. This shows that these are extensive quantities as one would expect it for the number of particles.

To compare systems of different sizes we will divide these quantities by the number of modes $N_{\text{mod}}$, i.e. the number of possible $\vec k$ values, with $\tilde n =  \hat n/ N_{\text{mod}} $. The expectation value $\left<\tilde n\right>$ can then be seen as the average occupation number and the variance $\langle (\Delta \hat n)^2\rangle/N_{\text{mod}}$ as average variance per mode. Our system is fully characterized by the ratio $|J|/K$. This characterization can be seen from Eq.~\eqref{eq:DefUV} and  the connection of $r_{\vec k}$ to the Bogoliubov coefficients below Eq.~\eqref{eq:DefSquOp}. Therefore, if we decide on a certain type of lattice, the ratio of $|J|$ and $K$ determines the number of sublattice magnons which are involved in the building of the squeezed states.

\begin{figure}
		\begin{center}
			\includegraphics[width = 8.6cm]{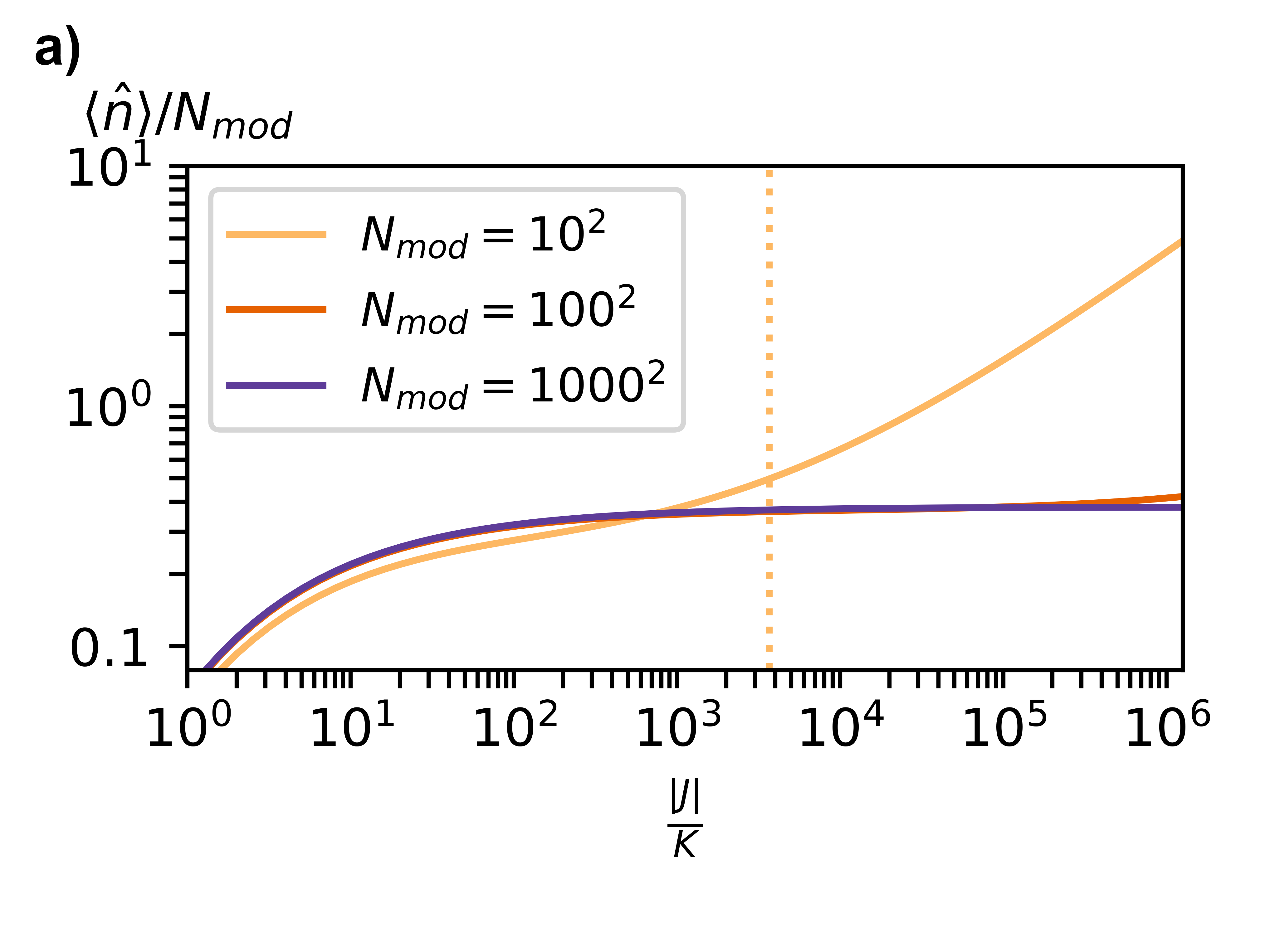}
		\end{center}
		\begin{center}
			\includegraphics[width = 8.6cm]{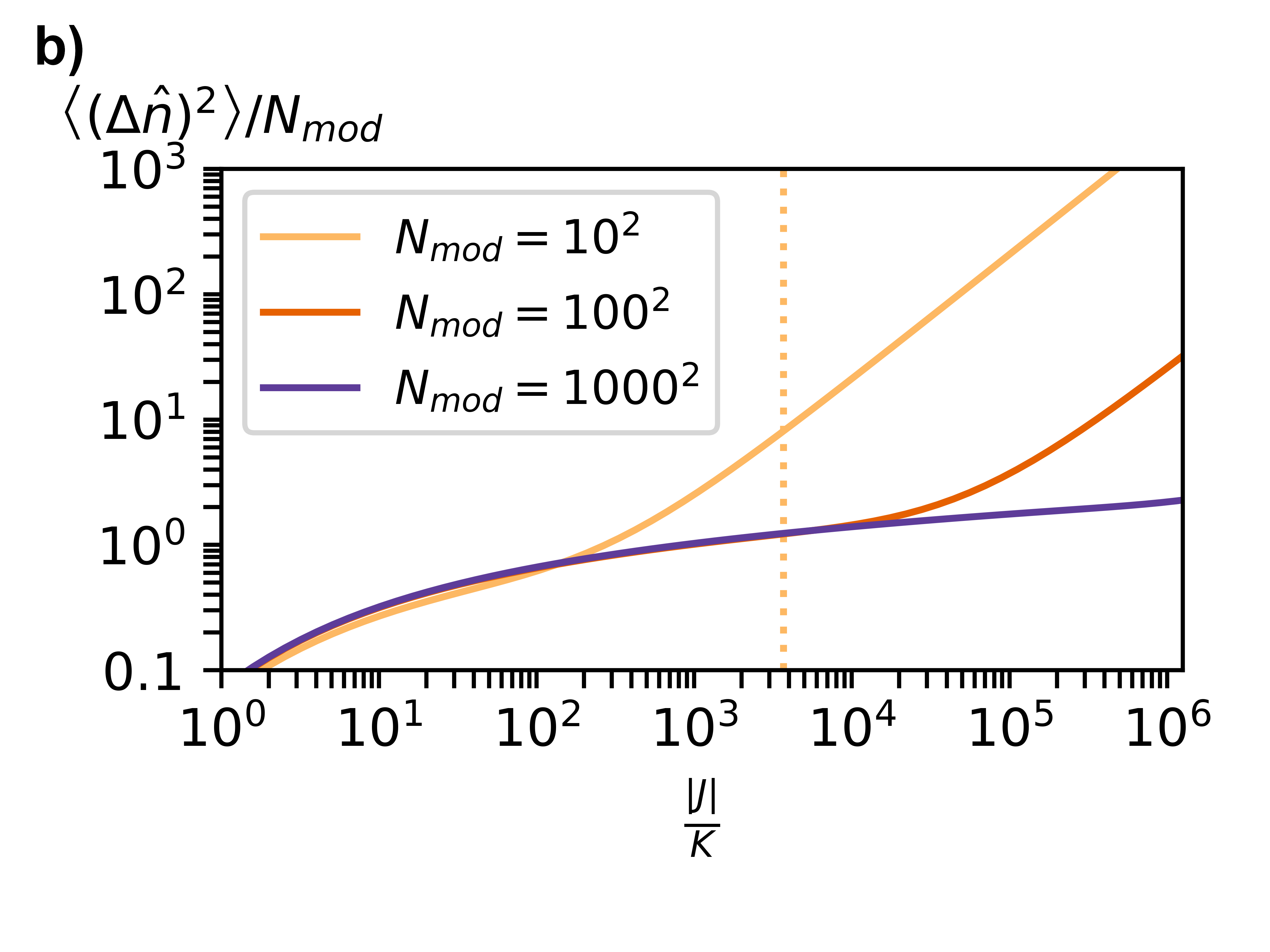}
		\end{center}
	\caption{Double logarithmic plot of $\left<\tilde n\right>$ (Fig.~a) and $\sqbraket{\left(\Delta \hat n\right)^2}/N_{\text{mod}}$ (Fig.~b) depending on $|J|/K$. The number of total modes $N_{\text{mod}}= N^2$, with $N$ being the number of sites in one dimension, varies from graph to graph.  The vertical, dotted line in a) shows the value for $|J|/K$, at which the $\vec k = \vec 0$ contribution to the total magnon number is as big as the total contribution from all other modes. This indicates the start of the linear behaviour for the magnon number in the double logarithmic plot.}
	\label{fig:4}
\end{figure}
\begin{figure}
		\begin{center}
			\includegraphics[width = 8.6cm]{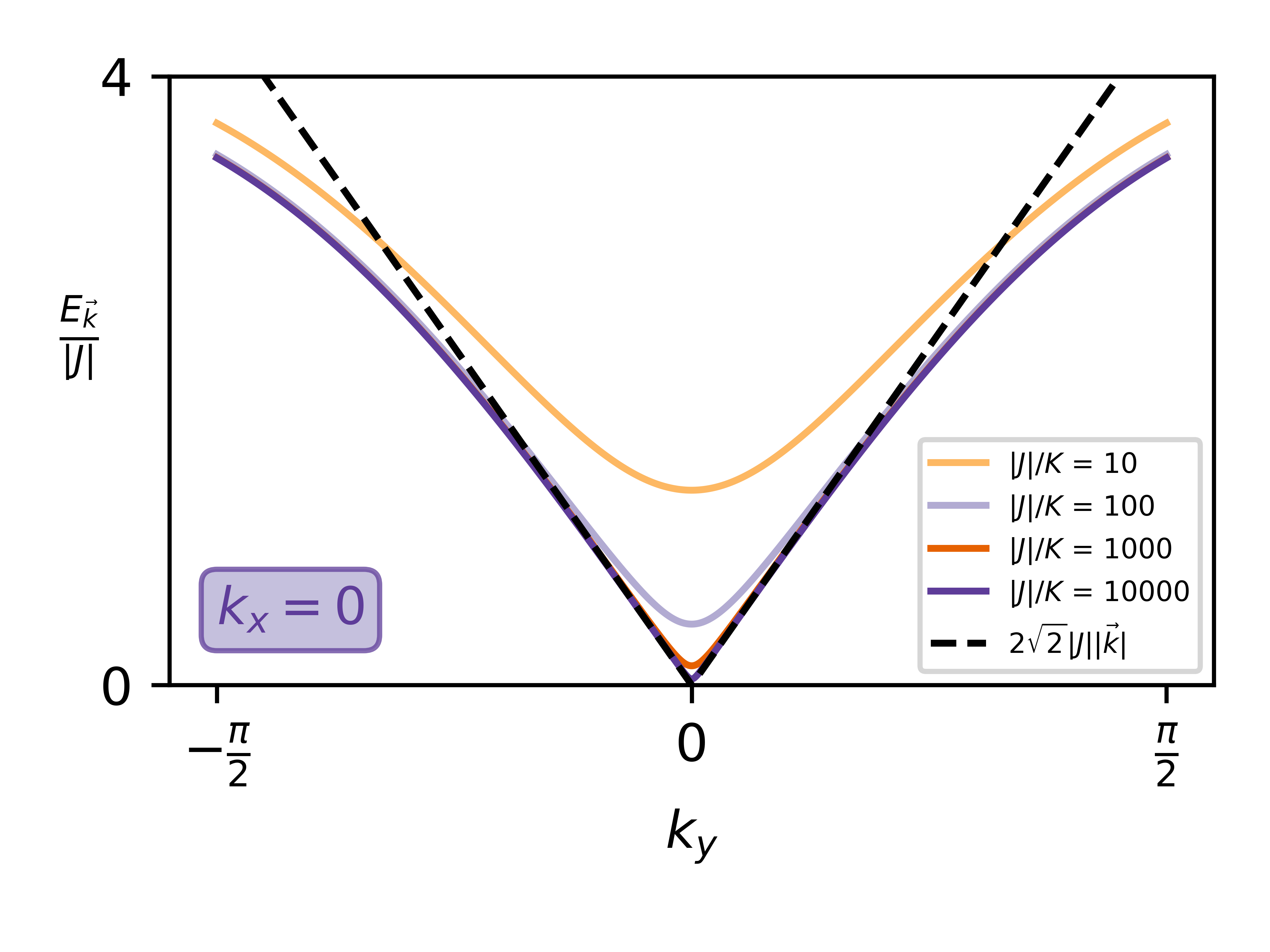}
		\end{center}
	\caption{Ratio of energy and $|J|$ for different values of $|J|/K$ and $J = -1$ showing the lowering of the energy gap for smaller $K$ and the approach to a linear dispersion. Dashed black line  gives the linear dispersion with $\varepsilon_{\vec k}/|J| = 2\sqrt{2}|\vec k|$.}
	\label{fig:5}
\end{figure}
Fig.~\ref{fig:4} shows the dependence of $\langle\tilde n\rangle$ and $\langle (\Delta \hat n)^2\rangle/N_{\text{mod}}$ on $|J|/K$ in a double logarithmic plot. The different graphs are given for distinct system sizes, i.e. distinct number of sites $N$ in one dimension. For a two dimensional square lattice, as it was assumed in Fig~\ref{fig:4}, one gets $N_{\text{mod}} = N^2$.  Even for different system sizes, one can see a few common features, e.g. the overall form of $\left<\tilde n\right>$ with a saturation around $\left<\tilde n\right> \approx 0.35$. As this can be interpreted as the average occupation number of the different modes, we can assume the linear approximation of the Holstein-Primakoff transformation, which demands a small number of magnons per mode, to hold, especially for larger spin lengths $S$. Only in the case of small samples one gets a strong deviation. This is reasonable as we expect finite size effects to become important and the approximation of periodic boundary condition becomes more and more inaccurate. Further the huge number of magnons implies, that in this case the linear spin wave approximation breaks down and higher terms in the creation and annihilation operators should be considered.

To estimate a value of $|J|/K$ at which we expect the approximation to break down, which is estimated for the $10\times10$ sublattice by the dashed line in Fig.~\ref{fig:4} a), we regard the large $|J|/K$ limit of the expectation value of the magnon numbers for different $\vec k$ values
\begin{equation}
		\braket{\hat n_{\vec k}}  \overset{\frac{|J|}{K}\gg 1}{=} \left\{\begin{array}{cc}
		\frac 1 2 \sqrt{\frac{|J|}{K}} & \qquad \left|\vec k\right| \leq \sqrt{\frac {2K} {\left|J\right|}}\\
		&\\
		\frac{1}{2\sqrt{1 - \gamma_{\vec k}^2}}-\frac 1 2 & \qquad \left|\vec k\right| \geq \sqrt{\frac {2K} {\left|J\right|}}
		\end{array}\right. \quad 
		\label{eq:ModeOccNumb}
\end{equation}
The square root dependency comes from the modes with  $|\vec k| \leq \sqrt{ 2K/ |J|}$, which goes towards the sole mode of $\vec k = \vec 0$ for $|J|/K \to \infty$. This is the same point at which a linear dependence of the energy on the absolute value of the wave vector is to be expected, as the contribution of the wave vector dependent part begins to dominate the energy. See therefore App.~\ref{SecA2}. For modes with $|\vec k|$ bigger than this value the contribution saturates at a, with respect to $|J|/K$, constant value. 

From Eq.~\eqref{eq:ModeOccNumb} it becomes clear, that the increase of the average magnon number for large $J/K$ in Fig.~\ref{fig:4} and, therefore, a breakdown of the linear spin wave theory, will be dominated especially by the $\vec k = 0$ magnon mode. The linear spin wave approximation demands $\langle\hat n_i\rangle \ll 1$. Clearly this condition will be violated if the average magnon number diverges. To get a quantitative value $|J|/K$ at which this approximation becomes invalid, we may have a look at the dominant $\vec k = 0$ mode that has the highest occupation number  and the linear spin wave approximation will be invalid for this mode first. 
The condition $\langle\hat n_i\rangle \ll 1$ in Fourier space implies for the $\vec k = 0$ mode that $\langle\hat n_0\rangle \ll N_{\text{mod}}$. Using Eq.~\eqref{eq:ModeOccNumb} we can calculate the approximate breakdown of the theory at $|J|/K \approx N_{\text{mod}}^2$. This is clearly confirmed by the data shown in Fig.~\ref{fig:4} for different systems sizes. 

From Eq.~\eqref{eq:VarianceN} it can be seen, that the variance, if plotted double logarithmically, also starts to grow linearly in $|J|/K$ for large values. The slope will be two times as big as for the expectation value. This implies, that even if the expectation value of the average occupation number grows, its variance grows even faster. 

From a physical point of view the divergence of the magnon number, arises, because the energy gap $\Delta$ goes towards zero for $K\rightarrow 0$(see Fig.~\ref{fig:5}). Therefore, as magnons are bosons, a huge number of magnons would be created, that populate the lowest energy level ($\vec k = \vec 0$), similar to the Bose-Einstein condensation studied in \cite{BoseEinsteinMagnon}. Again, the linearized spin wave theory obviously fails in this regime. On the other side, if $K$ becomes very large, i.e. $|J|/K \rightarrow 0$, we find a vanishing magnon number. This is obviously due to the fact, that the lowest magnon energy becomes huge and the spins are fixed to the anisotropy direction.

\section{Conclusion}
We picked up the work from Kamra et al.\ \cite{SqueezedMagnons} and further included all possible sublattice magnon states into the expansion of the squeezed magnon states. We were able to show that the probability amplitudes take the same form for each $\vec k$-component only differing in the squeezing parameter $r_{\vec k}$. 

From the squeezing parameter we determined the $k$-space structure of the probability to find at least one pair of sublattice magnons, which resulted in a $k$-space structure of the correlators of the magnetization and the N\'eel vector components. This enables experimental access to the probability structure by measurement of these correlators.
Further we also determined the correlators between the $x$-component of the magnetization  and the $y$-component of the N\'eel vector, which is purely imaginary and vanishes if we regard a symmetrized version, by using the anticommutator of the $x$-component of the magnetization  and the $y$-component of the N\'eel vector.

We determined the expectation value and the variance of the occupation number of each magnon mode each sublattice and from this we obtained the  $|J|/K$ dependence of the average occupation number of magnons in the system. We found that the average occupation number behaves similarly for different system sizes. From the expectation value of the number of sublattice magnons it is possible to identify the strength of the influence of different interactions on our Heisenberg AFM by their coupling either to sublattice magnons or squeezed magnons. 

Further investigations could concentrate on the effect of an applied magnetic field on the number of sublattice magnons participating in the squeezed Fock state. For a static magnetic field we expect a change in the number of $\vec k = \vec 0$ magnons, which will result in a simultaneous tilt of all sublattice spins. 

\textit{Post completion note:} After finishing this work, but before publishing, the authors became aware of the recent publication of Mousolou et. al.\cite{Mousolou}, which has a certain overlap with topics discussed above. While there are common themes, like the parameterization of relevant quantities in terms of the squeezing parameter $r_{\vec k}$, there are also differences, as the main focus on the entanglement entropy and a suggestion for an experimental setup in Mousolou et. al. work, while our publication is focused on the correlators of the magnetization and the N\'eel vector as well as the occupation number of magnons and their variance. We therefore see both works as complementary to each other.   

\section*{Acknowledgment} We acknowledge useful discussions with Akashdeep Kamra. This work was financially suppported by the Deutsche Forschungsgemeinschaft (DFG, German Research Foundation) via the Collaborative Research Center SFB 1432 (project no.~425217212) and via the Priority Program SPP 2244 (project no.~443404566). NR acknowledges financial support by the DFG via the project no.~417034116.

\appendix
    \section{Expansion of squeezed states in sublattice states }
	We start the expansion of the squeezed vacuum state in the sublattice states by acting with $\hat \alpha_{\vec k}$ onto the squeezed vacuum state and use the Bogoliubov transformation to determine the expansion coefficients
	\begin{align}
			\hat \alpha_{\vec k} \sqket{0} &=  \sum_{\vec n,\vec m = 0}^{\infty} A_{\vec n, \vec m} \left(u_{\vec k}\hat a_{\vec k} - v_{\vec k}\hat b_{-\vec k}^\dagger \right)\subket{\vec n ,\vec m} = \notag\\
			& = 0 \, .
	\end{align} 
	Here $\vec n$ ($\vec m$) contains the number of sublattice magnons with certain wave vectors in the A (B) sublattice. Doing the same with $\hat \beta_{\vec k}$ results in the following two equations
	\begin{align}
		A_{\vec n, \vec m} &= - \tanh(r_{\vec k})\sqrt{\frac{m_{-\vec k} }{n_{\vec k}}} A_{\vec n - \vec e_{\vec k} , \vec m - \vec e_{-\vec k}} \, ,\\
		A_{\vec n, \vec m} &= - \tanh(r_{\vec k})\sqrt{\frac{n_{-\vec k} }{m_{\vec k}}} A_{\vec n - \vec e_{-\vec k} , \vec m - \vec e_{\vec k}} \, .
	\end{align}
	These two equations yield, that the number of magnon with opposite wave vector in each sublattice is the same and a recursion relation, which lets us determine the factors $A_{\vec n, \vec m}$ in terms of the lowest coefficient $A_{0,0}$. $A_{0,0}$ can then be determined by the normalization condition of the squeezed vacuum. This all together yields
	\begin{equation}
		A_{\vec n, \vec m} = A_{\vec n} = \prod_{\vec k} \left[\frac{\left(-\tanh(r_{\vec k})\right)^{n_{\vec k}}}{\cosh(r_{\vec k})}\right]\,.
	\end{equation}
	\section{Linearity of the energy dispersion relation}
	\label{SecA2}
	For Fig.~\ref{fig:3} b) we limited the wave vector components on a value $k_0$. In Eq.~\eqref{eq:ModeOccNumb} we referred to a value for the absolute value of $\vec k$ for the conditional occupation number. Both values are the same and stem from the following analysis. If one assumes small values for the wave vector $\vec k$ in $N$ dimensions, one can approximate 
	\begin{equation}
	\sum_{\alpha}\cos{\left(k_\alpha\right)} \approx  N - \frac {\vec k ^2} 2 \qquad \qquad \alpha \in \left\{x_1, \ldots , x_N\right\}\, .
	\end{equation}
	Regarding a square lattice with nearest-neighbour interaction, we know that the dimension $N$ is half the number of nearest neighbours $z$ of one lattice side and is set by the dimensionality of the system. In the energy dispersion relation (Eq.~\eqref{eq:DefUV}) this yields
	\begin{equation}
	    \varepsilon_{\vec k} = \sqrt{A^2 - C_{0}^2\left( 1 - \frac{2 \vec k^2}{z}\right)} \,.
	\end{equation}
	We now determine the value for $|\vec k_0|$, by demanding, that the term proportional to $\vec k^2$ is as large as the other terms. This yields
	\begin{align}
	    \left|\vec k_0\right| &= \sqrt{\frac z 2 \frac{A^2 - C_0^2}{C_0^2}} = \sqrt{\frac z 2\frac{\left(1 - \frac{Jz}{K}\right)}{\left(\frac{Jz}{2K}\right)^2}}\\ &\overset{|J|/K \gg 1}{=} \sqrt{\frac{2 K}{|J|}}\, .
	\end{align}
	For wave vectors around this point we get a linear dispersion relation for the energy, as we can approximate 
	\begin{align}
	    \varepsilon_{\vec k} &\approx \sqrt {2\left(A^2 - C_0^2\right)} + \sqrt{\frac 2 z C_0^2}\left(\left|\vec k\right| - \left|\vec k_0\right|\right)\\
	    &=\sqrt {A^2 - C_0^2} + \sqrt{\frac 2 z C_0^2}\left|\vec k\right|\, .\notag
	\end{align}
	\section{Correlators}
	\label{SecA3}
	We want to give a short derivation of the correlators. starting from Eq.~(\ref{eq:MagNelk}) we define
	\begin{equation}
		\vec S_{\vec k} = \vec S_{\vec k}^A + \vec S_{\vec k}^B \, ,\qquad  \vec Q = \vec S_{\vec k}^A - \vec S_{\vec k}^B \, .
	\end{equation}
	The magnetization (N\'eel vector) is then given as \mbox{$\vec M_{\vec k} = \gamma \vec S_{\vec k}$} ($\vec N_{\vec k} = \gamma \vec Q_{\vec k}$). 
	\subsubsection{x-y-Components}
	We are interested in the correlators of the ladder components $\hat S^\pm_{\vec k}$ and $\hat Q^\pm_{\vec k}$ as they determine the correlators of the x and y components.	Due to the definition of the Fourier components of the ladder operators we get
	\begin{equation}
		\begin{split}
			\hat S_{\vec k}^{A,+} &= \sqrt{\frac 1 { N_{\text{mod}}}}\sum_{\vec r\in A} e^{\mathrm i \vec k \vec r}\sqrt{2S\hbar^2}\hat a\left(\vec r\right) = \sqrt{2S\hbar^2}\hat a_{\vec k}\, ,\\
			\hat S_{\vec k}^{A,-} &= \sqrt{\frac 2 { N_{\text{mod}}}}\sum_{\vec r\in A} e^{\mathrm i \vec k \vec r}\sqrt{2S\hbar^2}\hat a^\dagger\left(\vec r\right) = \sqrt{2S\hbar^2}\hat a^\dagger_{-\vec k}\, ,
		\end{split}
	\end{equation}
	with a similar expression for the B sublattice. ${ N_{\text{mod}}}$ is the number of modes in the system, which is half the number of sites, as we have two different sublattices. We get for $\hat S^+_{\vec k}$ 
	\begin{align}
		\begin{split}
		\hat S^+_{\vec k} &=S^{A,+}_{\vec k}+S^{B,+}_{\vec k}= \sqrt{2S\hbar^2} \left(\hat a_{\vec k} + \hat b_{-\vec k}^\dagger\right)=\\&= \hbar \sqrt{2S}\left[\left(u_{\vec k} + v_{\vec k}^*\right)\hat \alpha_{k} + \left(u_{\vec k}^\ast + v_{\vec k}\right)\hat \beta_{-k}^\dagger\right]	\\
		&= \left(\hat S^-_{-\vec k}\right)^\dagger \, ,
		\end{split}\\
		\begin{split}
			\hat Q^+_{\vec k} &=S^{A,+}_{\vec k}-S^{B,+}_{\vec k}= \sqrt{2S\hbar^2} \left(\hat a_{\vec k} - \hat b_{-\vec k}^\dagger\right)=\\&= \hbar \sqrt{2S}\left[\left(u_{\vec k} - v_{\vec k}^*\right)\hat \alpha_{k} -\left(u_{\vec k}^\ast - v_{\vec k}\right)\hat \beta_{-k}^\dagger\right]	\\
			&= \left(\hat Q^-_{-\vec k}\right)^\dagger \, .
		\end{split}
	\end{align} 
	Now one calculates the different combinations of products, i.e. $++$, $+-$, $-+$ and $--$, and takes the expectation value with respect to the squeezed vacuum. Only terms which contain the same number of creation and annihilation operators of the same kind of magnons can contribute. Further, as we are in the vacuum of the squeezed magnon modes, only modes with all creation operators on the right and all annihilation operator on the left are non-zero. This reduces the expectation values to
	\begin{align}
		\sqbraket{\hat{S}_{\vec k}^+\hat S_{\vec q}^-} &= 2S \hbar^2 \left(u_{\vec k} + v_{\vec k}^*\right)\left(u_{\vec q}^\ast + v_{\vec q}\right)\sqbraket{\hat \alpha_{\vec k}\hat \alpha_{-\vec q}^\dagger} = \notag\\
		& = 2 S \hbar^2\left|u_{\vec k} + v_{\vec k}^*\right|^2 \delta_{\vec k,- \vec q}\, ,\\
		\sqbraket{\hat{Q}_{\vec k}^+\hat Q_{\vec q}^-} &= 2S \hbar^2 \left(u_{\vec k} - v_{\vec k}^*\right)\left(u_{\vec q}^\ast - v_{\vec q}\right)\sqbraket{\hat \alpha_{\vec k}\hat \alpha_{-\vec q}^\dagger} = \notag\\
		& = 2 S \hbar^2\left|u_{\vec k} - v_{\vec k}^*\right|^2 \delta_{\vec k,- \vec q} \, ,\\
	    \sqbraket{\hat{S}_{\vec k}^+\hat Q_{\vec q}^-} &= 2S \hbar^2 \left(u_{\vec k} + v_{\vec k}^*\right)\left(u_{\vec q}^\ast - v_{\vec q}\right)\sqbraket{\hat \alpha_{\vec k}\hat \alpha_{-\vec q}^\dagger} = \notag\\
		& = 2 S \hbar^2\left(\left|u_{\vec k}\right|^2 - \left|v_{\vec k}\right|^2 \right)\delta_{\vec k,- \vec q} \,.
	\end{align}
	In the last line we dropped the resulting $\text{Im}(u_{\vec k}v_{\vec k})$ part, because $u_{\vec k}$ and $v_{\vec k}$ are real, as can be seen in the main text.
	
	All other combinations of ladder operators can be calculated in a similar manner. The combination of the $-+$ operators yield the exact same result (except for the $-+$ crosscorrelator, which gets a minus sign), while the combination of the same operators ($++$ and $--$) are always equal to zero due to the absence of terms with the same number of annihilation and creation operators of the same mode. 
	
	From the expectation values of the $+-$ components we can determine the expectation values of $xx$, $yy$ and $xy$ combinations due to their relation to the ladder operator components via 
	\begin{equation}
		\hat S_{\vec k}^x = \frac 1 2 \left(\hat S_{\vec k}^+ +  \hat S_{\vec k}^-\right)\, , \qquad 	\hat S_{\vec k}^y = \frac 1 {2 \mathrm i} \left(\hat S_{\vec k}^+ -  \hat S_{\vec k}^-\right)\, .
	\end{equation}
	Again, a similar relation holds for the components of $\vec Q_{\vec k}$. From this expression we can calculate the correlators to be equal to the values in Eqs.~(\ref{eq:CorrMag}), ~(\ref{eq:CorrNel}). The second part in the definition of the correlators will vanish, due to the fact, that each component is linear in the creation and annihilation operator of the magnons and therefore their expectation value will be zero. 
	\subsubsection{z-Components}
	Left to show is, that the correlators including z-components will vanish. From the definition we get 
	\begin{align}
		\hat S^z_{\vec k} &= \hbar \sqrt{\frac 1 { N_{\text{mod}}}}\sum_{\vec k_1}\left(\hat b_{\vec k_1} ^\dagger\hat b_{\vec k_1 + \vec k} - \hat a_{\vec k_1}^\dagger \hat a_{\vec k_1 + \vec k} \right)\, ,\\
		\begin{split}
		    \hat Q^z_{\vec k} &=2 \hbar S \sqrt{{ N_{\text{mod}}}}\delta_{\vec k, 0} - \\&-\hbar \sqrt{\frac 1 { N_{\text{mod}}}}\sum_{\vec k_1}\left(\hat b_{\vec k_1} ^\dagger\hat b_{\vec k_1 + \vec k} + \hat a_{\vec k_1}^\dagger \hat a_{\vec k_1 + \vec k} \right)\, .
		\end{split}
	\end{align}
	As we only regard term up to order two in the annihilation and creation operators, it is obvious, that all products containing the z-component of the magnetization will have a zero expectation value. For the expectation value of each z-component we have to consider 
	\begin{equation}
		\begin{split}
		\sqbraket{\sum_{\vec k_1}a_{\vec k_1}^\dagger \hat a_{\vec k_1 + \vec k}} &=\sum_{\vec k_1}\left|v_{\vec k_1}\right|^2\delta_{\vec k,0}\\&= \sqbraket{\sum_{\vec k_1}b_{\vec k_1}^\dagger \hat b_{\vec k_1 + \vec k}} \, ,
	\end{split}
	\end{equation}
	which implies $\sqbraket{\hat S^z_{\vec k}}= 0$ and  
	\begin{equation}
		\sqbraket{\hat Q^z_{\vec k}} = 2\hbar\left(  S \sqrt{N_{\text{mod}}}-  \sqrt{\frac 1 {N_{\text{mod}}}}\sum_{\vec k_1}\left|v_{\vec k_1}\right|^2\right)\delta_{\vec k, 0} \, .
	\end{equation}
	From this it is a short calculation to show, that the correlator of the z-components of the N\'eel vector and of the z-components of the magnetization and the N\'eel vector vanishes. Further all products of other components (odd number of operators) with the z-components of the N\'eel vector (even number of operators) will also vanish if we take the expectation value. 
	\newpage

\bibliography{MagNumBib}

\end{document}